\documentclass[conference]{IEEEtran}

\usepackage{graphicx}
\usepackage{algorithm}
\usepackage{algorithmic}
\usepackage{booktabs}
\usepackage{textcomp}
\usepackage{makecell}
\usepackage{cite}
\usepackage{url}

\begin{document}

\newcommand{\tabincell}[2]{\begin{tabular}{@{}#1@{}}#2\end{tabular}}

\title{Rule Optimization for Real-Time Query Service in Software-Defined Internet of Vehicles}

\author{\authorblockN{Xin Wang\authorrefmark{1}, Cheng Wang\authorrefmark{1}, Changjun Jiang\authorrefmark{1}, Lei Yang\authorrefmark{2}, Zhong Li\authorrefmark{1}, Xiaobo Zhou\authorrefmark{3}}
\authorblockA{\authorrefmark{1} Department of Computer
Science, Tongji University,  China}
\authorblockA{\authorrefmark{2}  Institute of Trustworthy Networks and Systems, Tsinghua University,  China}
\authorblockA{\authorrefmark{3}  Department of Computer Science, University of Colorado, USA}
}

\maketitle

\begin{abstract}
Internet of Vehicles (IoV) has recently gained considerable attentions from both industry and research communities since the development of communication technology and smart city. However, a proprietary and closed way of operating hardwares in network equipments slows down the progress of new services deployment and extension in IoV. Moreover, the tightly coupled control and data planes in traditional networks significantly increase the complexity and cost of network management. By proposing a novel architecture, called Software-Defined Internet of Vehicles (SDIV), we adopt the software-defined network (SDN) architecture to address these problems by leveraging its separation of the control plane from the data plane and a uniform way to configure heterogeneous switches. However, the characteristics of IoV introduce the very challenges in rule installation due to the limited size of Flow Tables at OpenFlow-enabled switches which are the main component of SDN. It is necessary to build compact Flow Tables for the scalability of IoV. Accordingly, we develop a rule optimization approach for real-time query service in SDIV. Specifically, we separate wired data plane from wireless data plane and use multicast address in wireless data plane. Furthermore, we introduce a destination-driven model in wired data plane for reducing the number of rules at switches. Experiments show that our rule optimization strategy reduces the number of rules while keeping the performance of data transmission.
\end{abstract}


\section{Introduction}

Internet of Vehicles (IoV) is attracting considerable attention from both academia and industry. The vigorous development of communication technology and smart city makes various services possible in IoV, which significantly improves the quality and safety of driving. The research and industry communities are carrying out several projects\cite{CVIS}\cite{makino2005smartway} for the development of IoV. For example, EU's CVIS \cite{CVIS} is committed to design, develop and test the technologies needed to allow cars to communicate with each other and with the nearby roadside infrastructure. Smartway \cite{makino2005smartway} focuses on integrating all ITS functions to a Smartway platform and provides services by two-way communication.

Though there is a promising future in IoV, the proprietary and closed way of operating hardwares in network equipments slows down the progress of new services deployment and extension in IoV. Network equipments such as switches and routers are developed by different manufacturers. Every change of network equipments requires substantial manual configuration by trained operators, which makes network management expensive and error-prone. The lack of an open and unified interface for flexible and dynamically customizable network makes new services deployment and extension difficult in large-scale IoV. A new network architecture is expected for the development of IoV.

\begin{figure} [t]
\begin{center}
\begin{tabular}{cc}
\includegraphics[width=0.48\columnwidth]{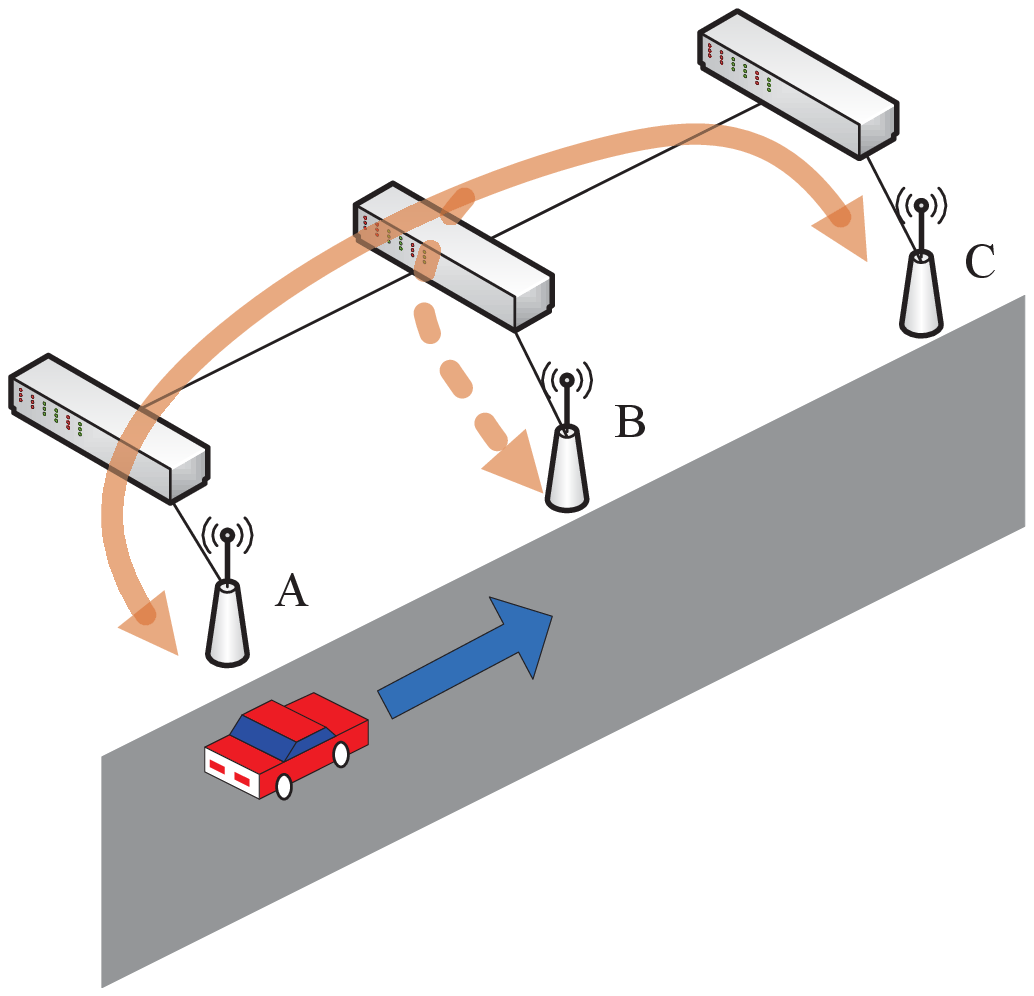}&
\includegraphics[width=0.48\columnwidth]{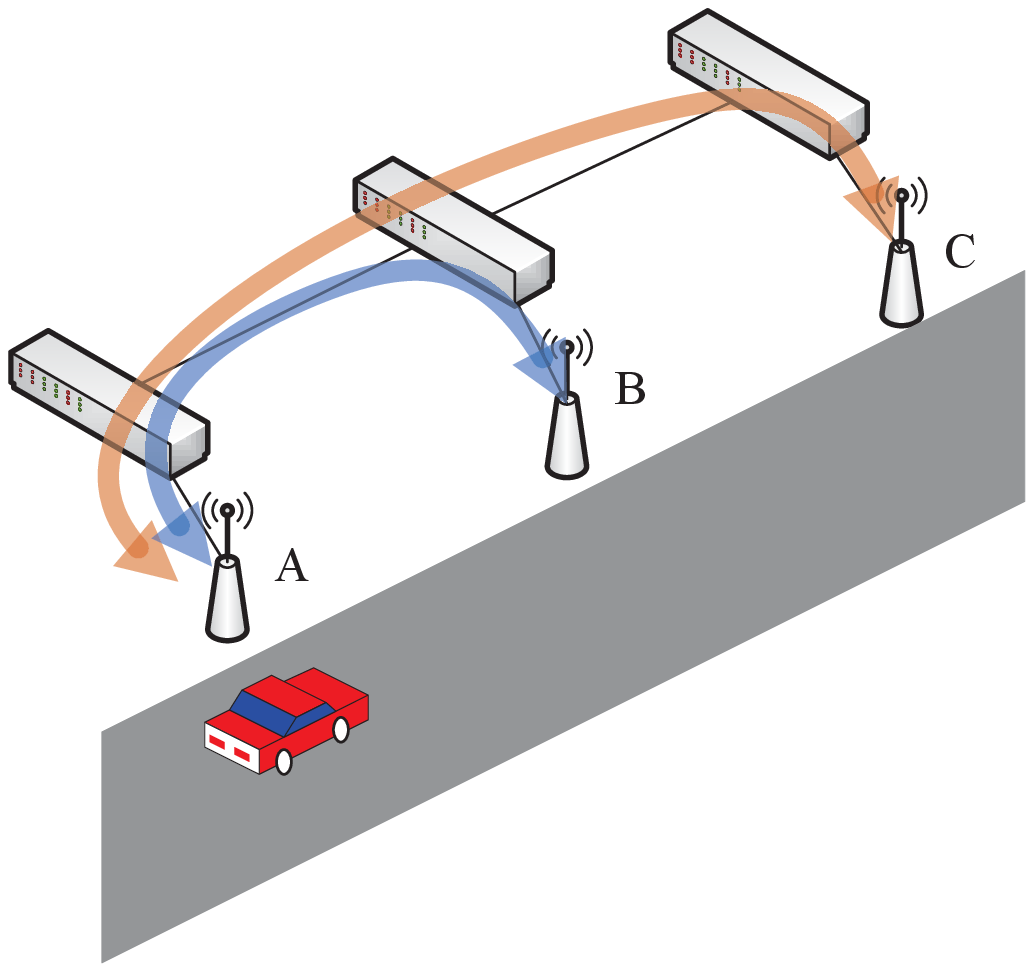} \\
(a) & (b)
\end{tabular}
\caption{In Subfigure (a), when the vehicle moves to another place (form $A$ to $B$), it needs to establish a new connection. The dash line represents the connection which is not established yet; In Subfigure (b), when a vehicle connecting to multiple cameras simultaneously, it needs to build a path for each connection which is not efficient since all paths have the same destination (all data flow's destination is $A$).} \label{fig1}
  \end{center}
  \vspace{-0.3in}
\end{figure}

There are several projects for the new network architecture and Next Generation Internet (NGI). Named Data Networking (NDN) \cite{NDN} aims to develop a new Internet architecture that concentrates on getting `what' service rather than `where' to get service. MobilityFirst \cite{MobilityFirst} supports seamless and smooth mobility and it takes the mobility of nodes as a common case rather than a special case in traditional networks. NEBULA \cite{NEBULA} develops a new network architecture based on cloud computing and data centers. The eXpressive Internet Architecture (XIA) \cite{XIA} supports the trustworthy communication, the growing diversity of network using models, the sustained technological innovation and clarity of the interface between different networks. Software-Defined Network (SDN) provides a unified interface to configure network equipments and separates the control plane from the data plane. A unified interface of configuring network equipments makes large scale customizable network possible, and accelerates new services deployment in IoV. Therefore, we adopt SDN to support IoV for its open and unified interface, and propose Software-Defined Internet of Vehicles (SDIV), a new architecture for the development of IoV. SDN has several advantages in supporting IoV besides the open and unified interface: (1) SDN naturally has a high scalability by separating the data plane from control plane, (2) SDN has good network management capabilities as it centralizes the control part to the controller, (3) the controller is able to choose the best path for data transmission and plays as a coordinator between roadside electronic devices according to current network state. But, the characteristics of IoV introduce challenges in rule installation due to the limited size of Flow Tables at OpenFlow switches. It is necessary to build compact Flow Tables for scalability of IoV.

In this paper, we consider the real-time query service, one of typical services in IoV, to show the design details for rule optimization in SDIV. Real-time query service provides drivers the real-time conditions of roads and then drivers decide which way to go according to the information. Road information coming from the surveillance cameras (or other kind of roadside electronic devices) will be transmitted to every vehicles with the information demanded. In this scenario, the specific issues are: (1) the mobility of vehicles increases the number of rules since it needs to establish new connections between vehicles and surveillance cameras, (2) when a vehicle connecting to multiple cameras, it is not efficient to install rules for every path since each path has the same destination. Figure 1 illustrates the problems of rule installation in IoV. If the controller simply installs rules upon the requests of drivers, the table size at switches will be the bottleneck of scalability, which impacts the performance of real-time query service.

To address the rule optimization problems for real-time query service, we introduce several techniques by leveraging the centralized and fine granularity of data flow control in SDN for rule optimization: (1) To reduce the number of requests sent by vehicles, we use multicast address instead of general destination address as the last address when vehicles receive data from cameras; (2) To keep uninterrupted connection between vehicles and roadside electronic devices, we make the controller install rules in advance based on the conditions of vehicles; (3) To decrease the number of rules and maintain the correctness of data transmission, we modify the headers when packets come to branching nodes.

In this work, we first describe the architecture of SDIV. We then analyze the problem of flow table size in OpenFlow supported switches and design and develop an approach to reduce the number of rules in switches. We validate the feasibility of rule optimization by considering four situations in the real-time query service and analyzing the details of data transmission in each case. Finally we use Floodlight \cite{Floodlight} as controller and Mininet \cite{mininet} as the testbed to evaluate the performance of our rule optimization method. Simulation results show using rule optimization, the flow table at switches have been more compact compared to simply installing rules and does not lose the performance.

The rest of the paper is organized as follows. In Section \ref{Architecture}, we introduce the architecture of SDIV. In Section \ref{Rule Optimization}, we discuss the rule optimization issues and show how it works. We conduct extensive experiments and report our results in Section \ref{Evaluation}.  In Section \ref{Related Work}, we review related works. We conclude the paper in Section \ref{Conclusion}.

\section{Network Architecture: Software-Defined IOV} \label{Architecture}

In this section, we propose Software Defined Internet of Vehicles (SDIV) architecture and discuss its utility through real-time query service in IoV. Before introducing the architecture of SDIV, we describe the basic SDN model and its main features.

\subsection{Software Defined Networks (SDN)}\label{SDN}

The principal endeavors of SDN are proposed to separate the control
plane from the data plane and centralize network's
intelligence and state to a single device. A SDN network consists of two parts: (1) switches process flows' packets based on actions of flow entries in the Flow Table, (2) the controller generally runs on a remote commodity server and communicates over a secure connection with the switches using a southbound interface to add and remove flow entries from the flow table. A flow table consists of flow entries associated with actions that tell the switch how to process the flow. Each flow entry has an action associated with it: (1) Forward this flow's packets to a given port so as to route packets through the network; (2) Encapsulate and forward this flow's packets to the controller for installing rules; (3) Drop this flow's packets. When a new switch joins the network, it will send a hello message to the controller. Then, the controller recognizes the switch and changes the network status appropriately. The controller interacts with the forwarding elements through the southbound interface. OpenFlow, a protocol maintained by ONF\cite{ONF}, can be viewed as a promising implementation of such
an interaction. OpenFlow enables users to control data flows in the network by installing rules with matching fields and actions of processing packets at switches. In this work, we use OpenFlow as the intermediate protocol between the controller and switches.

\subsection{Software-Defined IoV (SDIV)} \label{SDIV}

Our proposed SDIV network has a three-tier architecture. From bottom to top, they are physical layer, control layer and application layer, as illustrated in Figure 2.

\subsubsection{Physical Layer}

In physical layer, it is the same as IoV with vehicles, APs, roadside electronic devices, switches and servers. Vehicles act as mobile nodes and communicate with the server through road-side APs. When sensors in a vehicle collect information about conditions of the vehicle (e.g. speed, direction and location), the data should be transmitted to the server as soon as possible. And vehicles should receive responses from the server through nearby APs (we assume that the number of road-side APs is much enough to cover every road). Road-side APs are static WiFi access points reachable from the road that offer the ability of data transmission to vehicles. Roadside electronic devices like surveillance cameras gather road conditions and also send data to servers or vehicles. Switches connect APs, surveillance cameras and servers. The server analyzes the data gathered from APs and vehicles for information services, such as traffic conditions, local map, car accidents and provides services for user requests. Location information of units in IoV should be in the form of coordinates given in longitude and latitude, which can be easily found with GPS. Multiple servers are available for vehicles retrieving data from appropriate server considering the work load and location of both vehicles and the server.

\subsubsection{Control Layer}

In control layer, the controller connecting to every switches (including APs) acts as a coordinator for various services and data flows by installing rules at switches via OpenFlow Protocol. Switches forward any packet with no rule matched to the controller and then controller installs rules at the switches based on strategies pre-configured, and this enables the controller to control all data flows in IoV. When network status changed, the switches will notify the controller through OpenFlow Protocol. In this architecture, switches (including APs) act as connector between vehicles, servers and other roadside electronic devices by forwarding data flows based on rules. The controller has an up-to-date, global view of the network topology and traffic states, and provide the capability of customizing networks easily. Another important duty of the controller is abstracting the underlying network topology for the upper services, and providing network states to the applications in the upper layer. The abstract underlying topology should be suited to the application that each application has different purposes, in another word, different applications may have different views of the topology, and it does not need to provide whole network topology with all details to applications since they only care about a small part of the network topology. Also privacy and security are benefit from the abstract topology that enclosure the details of devices in the underlying network. The controller provides network states (i.e., link utility, the number of rules at switches and the number of flows at switches) to the applications in order to make the applications write their own strategies for their services implementation.

\subsubsection{Application Layer}

The strategy of each application is defined in the application layer. Applications provide services for drivers in IoV, such as real-time query service, location service and road conditions service. Each application get the network states from control layer and make decision according to their strategies. The strategy here means how to provide service defined in the application to the clients (e.g., the vehicles). Such as real-time query service is to provide drivers the real-time conditions of roads and then drivers decide which way to go according to the information, then the topology of the network is necessary to compute the path for improving performance which we will give details in Section \ref{Rule Optimization}. Applications also have abilities to install rules into the chosen switches by the interfaces provided from the control layer. The input of applications in this layer is the network states and the forwarded packets to invoke the application service, and the output of applications is the rules at selected switches.

\begin{figure} [t]
\begin{center}
\includegraphics[width=0.9\columnwidth,height=2.5in]{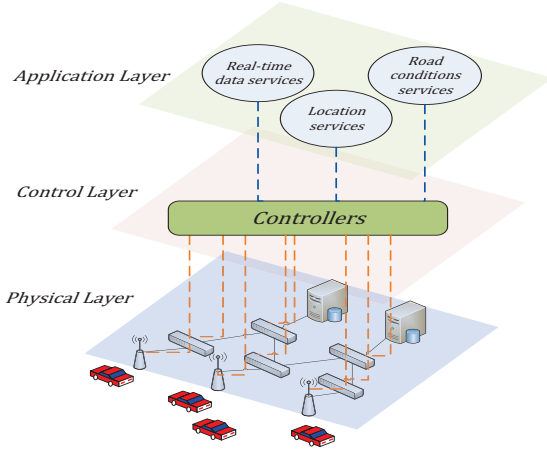}
\vspace{-0.2in}
\caption{Three-tier architecture of SDIV with physical layer, control layer and application layer bottom up.} \label{fig2}
\end{center}
\vspace{-0.3in}
\end{figure}

\subsection{SDIV operations and advantages} \label{sdiv oa}

After introducing the components of SDIV, we describe how SDIV works through two scenarios in SDIV as depicted in Figure 3.

\begin{figure} [t]
\begin{center}
\begin{tabular}{c}
\includegraphics[width=1\columnwidth]{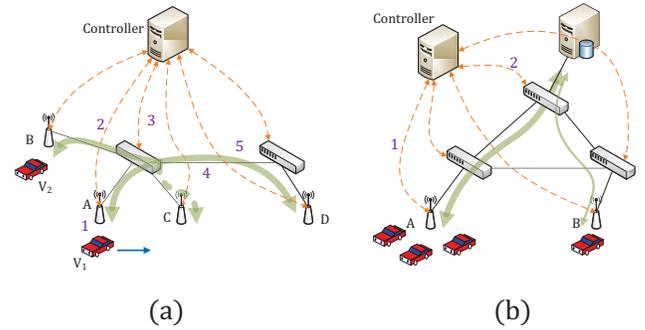} \\
\end{tabular}
\caption{Subfigure (a) shows a scenario that a vehicle want to get the information about road conditions maybe two or three blocks away from current location; Subfigure (b) depicts a scenario of data uploading through nearby AP to the server.} \label{fig3}
  \end{center}
  \vspace{-0.3in}
\end{figure}

Figure 3 (a) describes a typical scenario where a vehicle wants to get the information about road conditions (maybe three blocks away from current location) via surveillance cameras.

In Step 1, the vehicle $V_{1}$ sends a request to a nearby road-side AP as shown in Figure 3 (a); in Step 2, since there is no rule matching the header of the flow's first packet, the switch (AP $A$) encapsulates and forwards the packet to the controller; in Step 3, the controller recognizes the header and installs rules based on the vehicle's requests and information (e.g., location, speed and direction) and also current network's status; in Step 4, switches along the path towards the destination forward the matched packet to given ports based on the rules installed; in Step 5, the other side data flow, from the surveillance camera ($D$) to the vehicle, go through the same procedure as described in Step 2-4. When the number of vehicles increasing ($V_{2}$ at $B$ appears), there needs a scalable approach in installing rules for data transmission. The conditions (e.g., directions) of vehicles also should be considered for installing rules (install rules at $C$ in advance).

Figure 3 (b) describes a scenario where the vehicles need to build connections between the server through nearby AP for data uploading such as their vehicles conditions.

We can classify road-side APs by centrality degree (if an AP is connected by more vehicles, then it has a higher degree) in the controller which has a global view of the network. Traffic engineering can be implemented based on the level of services and the centrality degree of APs using LP \cite{danna2012practical} which is not concerned in this work. As illustrated in Figure 3 (b), initially vehicles send data to the server. Then, in Step 1, the controller gathers the information from AP $A$ and $B$ (when the switches or APs receive the packets with no rule matching, they will send the packets to the controller) for computing their centricity degrees. In Step 2, the controller installs rules at switches according to the centricity degree. In Figure 3 (b), AP $A$ has higher degree than AP $B$ since there are more vehicles connecting to $A$ than $B$, then we allocate more bandwidth for data flows coming from $A$.

In addition, SDIV is more attractive when considering large-scale multicast and the mobility of vehicles. As the number of vehicles connecting to one camera increasing, it is easy to think of multicast for the efficiency of data transmission. Although dense-mode multicast (reasonable for cameras) goes well in small networks, as the range of data transmission increasing, the growing number of $(S,G)$ entries kept in APs (or switches) and broadcasting messages periodically for establishing SPT limit the scalability of data transmission. The mixed model of control and data plane makes it necessary to keep $(S,G)$ entries even not used currently at every routers for a fast graft. In Figure 3 (a), it is necessary to reserve $(S,G)$ entries at AP $B$ for a new coming vehicle $V_{2}$. The mobility of vehicles also brings difficulty to traditional network technologies. Vehicle $V_{1}$ moving from $A$ to $C$ as described in Figure 3 (a) need to send another request for retrieving data, which results in interruption of data stream and moreover increases the work load of the surveillance camera ($D$). In our design, we address the mobility problem by predicting the most possible path that the vehicles would choose and set the rules in flow tables advance to multicast the packets from nearby APs along the path. Multicast along the data transferred path in SPT can not satisfy the mobility issue since the shortest path for data transmission seldom matches the driving path.

Table 1 summarizes the benefits of SDIV against traditional network technologies (multicast) besides its open interfaces of network equipments. Compared to traditional multicast method that needs to broadcast messages periodically for establishing SPT, SDIV leverages the benefits of OpenFlow protocol that makes switches forward packets with no matching rule in flow table to the controller and then the controller installs rules for the packets, which is reactive mode and it doesn't need to broadcast messages periodically. This reactive mode also makes is possible that the switches only need to keep the least number of rules in flow table against keeping $(S,G)$ entries at routers in traditional multicast method. With the support of the controller with an up-to-date, global view of the network topology and traffic states, it can compute the most possible path that the vehicles would choose and installs rules in advance for providing persistent connection between vehicles and the devices.

\begin{table}[t]
 \caption{\label{table1}Summary of benefits against traditional network technologies (multicast) in SDIV for location-based services}
 \centering
 \begin{tabular}{c|c}
  \hline
  Traditional Technologies & SDIV \\
  \hline
  \hline
 \Gape[4pt]Broadcasting messages periodically & Reactive mode \\
 \hline
 \Gape[4pt]Keeping $(S,G)$ entries at routers& Installing rules when needed \\
 \hline
 \Gape[4pt]SPT can not match the driving path & \tabincell{c}{Finding the path according to \\the direction of vehicles}\\
  \hline
 \end{tabular}
\end{table}

\section{Rule Optimization: Problem and Approach} \label{Rule Optimization}

Having described the overview architecture of SDIV and how SDIV works, we show more details about rule installation and explain that it could bring complexities if naively installing rules just for packets forwarding. In this section, we consider real-time query service as the context to explain the necessity of rule optimization.

In the real-time query service, if controller simply installs rules for the requests from drivers, the table size at switches will be the performance bottleneck since the limited size of flow table. When vehicles moving from one place to another, it needs to send another request packets which will be forwarded to the controller since there is no rule matching the packets. In this work, we assume that OpenFlow supported switches would select reactive mode that there is no rule at the beginning. The packets forwarded to the controller will lead to a serious computational bottleneck at the controller. Even if the controller's computational capacity can scale, the bandwidth demand that every packets go through the controller may be impractical. As for the drivers, the delay caused by every packets forwarded to the controller is much larger than the directly forwarding to the destination devices. Hence, it is very important to decrease the number of packets forwarded to the controller both for the controller's capacity and the drivers' using experience. Besides the number of packets forwarded to the controller, the number of rules at the switches is another factor that influences the performance of network since the limited size of TCAM at the switches, hence it is important to produce a compact table to fit more cached policy decision at the switch. For each flow established, it needs two rules at the selected switches for sending the request packets and receiving the data packets from the destination, and even if we only consider the second data flow, since the request flow is only useful at the beginning and can be deleted by timeout entry in OpenFlow switches, it still need one rule for each flow. It seems that one rule for each flow is compact enough for flow table, but in reality drivers always connect to multiple surveillance cameras at the same time and compare all roads' conditions then choose the best path. As a result, there needs to install rules for every data flow at every selected switches even though these flows all have the same destination. To summarize, we need to decrease the number of packets forwarded to the controller and establish a compact flow table since the limited size of TCAM at switches.

Figure 4 depicts the basic scenario that how real-time query service works. Suppose that vehicle $V_{1}$ wants to see the conditions of road (or crossroads) $E$. Here, the conditions can be represented as video or any other kinds of data stored in the roadside devices. At the first step, $V_{1}$ sends a request to AP $A$ and then the data flow is forwarded to $E$. After confirming the request from $V_{1}$, $E$ starts to send data to AP $A$. Finally, AP $A$ transmits the data to $V_{1}$ and completes the data transmission. The entire process is simple, but this simplicity may introduce performance lost if simply implemented. Consider the situation that when there is another vehicle $V_{2}$ also wants to get the information of $E$, as shown in Figure 4. $V_{2}$ will follow the same procedure as $V_{1}$ does, and there is another path from $E$ to $B$, which makes switch $D$ installed two rules following the same path. As a result, with the number of vehicles connecting to $E$ getting large, the size of flow table at switches along the path reaches the maximum, which reduces the performance of the service. Moreover, when $V_{1}$ moves to other place, it needs to send a request again.

Next, we consider another example to show the importance of rule optimization. $V_{1}$ wants to get the data from $F$ at the same time, then switch $B$ and $D$ need to install more rules to make data flow transmitted to $A$, even though both flows ($E \to A$ and $F \to A$) have a
same destination. Situations will become more complex if vehicle $V_{3}$ joins. To make $V_{3}$ at $C$ receive data, there needs to build a new path which increases table size. The mobility of vehicles increases the complexity and reduces performance as well.

To address the problem, we propose an idea to separate the wireless data plane (for communication between vehicles and APs) from wired data plane (for communication among switches) and develop a destination-driven model (the details of destination-driven model are shown in the Section \ref{Modify address}) for the wired data plane. When vehicles receive data from nearby APs, they do not care how data transmitted in wired data plane, they just want a persistent connection though their locations change over time. To make the separation efficiently, we introduce three techniques: (1) using multicast address as the last hop address from cameras to vehicles, (2) installing rules in the most possible path in advance according to the conditions of vehicles, (3) modifying the headers when packets come to branching nodes.

%

\subsection{Multicast Address for the Last Hop} \label{Multicast address}

We utilize multicast address from traditional networks for the mobility and scalability in SDIV. Every surveillance camera has a unique multicast address which can be generated from MAC address as the same method in traditional networks. Multicast address can be used in the last hop from cameras to vehicles for data transmission in wireless data plane. Vehicle $V_{1}$ receives packets with the destination address of $E$'s multicast address in Figure 4. The advantage of using multicast address is apparent that every vehicles, within the range of an AP which has been equipped the rule, can get the real-time data without sending new requests. Furthermore, in real-time query service, the first packet of users' requests have to be forwarded to the controller for installing rules. If every request needs to contact the controller for rule installation, it will lead to a serious bottleneck at the controller since the computational capacity and bandwidth are limited. A multicast address can address the problem efficiently. After the first vehicle connecting the camera with the intervention of the controller, the data flow will multicast around its location (along the path to the destination), and every new vehicle asking for the same camera can directly receive the data without sending the request, which reduces the frequency of contacting the controller. As the example in Figure 4, if there is another vehicle that wants to see the conditions of $E$, it can receive the data without sending a request after vehicle $V_{1}$ sending the request.

Due to the limited size of TCAM in OpenFlow supported switches, we need to remove the useless rules from flow tables. In our design, we use timeout in OpenFlow protocol to delete rules for saving flow table resources. A switch maintains a per-flow-entry variable that indicates if there has ever been a period of $T$ seconds in which no packet arrived for the flow. In practice in OpenFlow, this is maintained by adding an "idle timeout" value to a flow entry. Comparing to equal values of timeout, we choose selecting different values of timeout in switches depending on the distance away from the destination location. We make an assumption that the number of vehicles interested in the destination gets larger as they are closer to the destination, then we set larger values in timeout for the closer switches to the destination. As an example in Figure 4, the timeout value $T_{i}$ in switch $i$ has the constraint: $T_{A} < T_{B} < T_{D} < T_{E}$, and in our design, we set the value $k$ in the equation: $T_{i} = T_{j}*(k^{d_{ij}})$ where $d_{ij}$ means the distance between the destination $i$ and other location $j$. For simplicity, the distance can be the hop count.

\begin{figure} [t]
\begin{center}
\includegraphics[width=1\columnwidth]{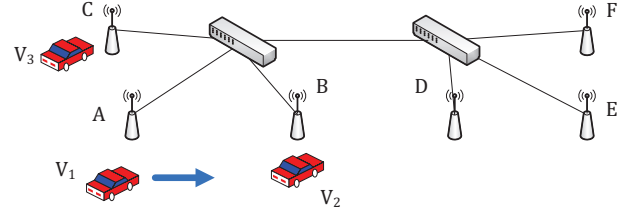}

\caption{$V_{1}$ is connecting to $E$ and $F$ simultaneously, and $V_{2}$ at the path of $V_{1}$ to $E$ asks the data from $E$. $V_{3}$ also requires the data from $E$ but at the different path.} \label{fig4}
\end{center}
\end{figure}

\subsection{Predict the Path and Install Rules in Advance} \label{Predict the path}

\begin{algorithm}[t]
\caption{PathFind($s,d,v$)}
\label{Algorithm 2}
\begin{algorithmic}[1]

\STATE{Algorithm PathFind($s,d,v$)}
\STATE{~~put $s$ into set $N$, $R$;}
\STATE{~~for each child node $c$ of $s$}
\STATE{~~~~if the angle between $v$ and $V(s,c)$ is less than 90 degree then}
\STATE{~~~~~~put $c$ into set $O$;}
\STATE{~~Find($s,O,d$);}
\STATE{~~return $R$;}
\STATE{~}
\STATE{Procedure Find($p,O,d$)}
\STATE{~~put $p$ into set $N$;}
\STATE{~~for each $n$ in $O$}
\STATE{~~~~if $n$ is $d$ then}
\STATE{~~~~~~put $n$ into set $R$;}
\STATE{~~~~~~finish;}
\STATE{~~~~put $n$ into set $N$;}
\STATE{~~~~remove $n$ from set $O$;}
\STATE{~~~~calculate $D(n,d)$, $D(p,n)$;}
\STATE{~~~~if $D(p,n)+D(n,d)<max$ then}
\STATE{~~~~~~$max = D(p,n)+D(n,d)$;}
\STATE{~~~~~~$r = n$;}
\STATE{~~put $r$ into set $R$;}
\STATE{~~for each child node $c$ of $r$}
\STATE{~~~~if $c$ not in set $N$ then}
\STATE{~~~~~~put $c$ into set $O$;}
\STATE{~~Find($r,O,d$);}
\STATE{~~return;}
\end{algorithmic}
\vspace{-0.06in}
\end{algorithm}

To deal with the mobility of vehicles in SDIV, it is necessary to predict the most possible path that a driver will choose, and then install rules in advance along the path in order to keep data transmission uninterrupted when the vehicle moves. A simple method to predict the path is computing the shortest path, but it is not always a correct path in reality. Now, we describe $PathFind(s,d,v)$, a simple algorithm shown in Algorithm 1 that finds the most likely path in topology. Here $s$ is the current location, $d$ is the destination and $v$ is the direction of the vehicle. $V(s,c)$ denotes the vector of $s$ and $c$, $D(n,d)$ means euclidean distance between $n$ and $d$.

As the preparatory work (Lines 3-5), $PathFind$ filters the possible first node of result path according to the condition (e.g., direction) of the vehicle, since the driver seldom turn back in street. Then, we apply $Find$ recursively to calculate the result path by choosing the node which has the minimum value of $D(p,n)+D(n,d)$ (e.g., the distance between the chosen location to destination) as shown in Lines 17-20. In current design, $D(n,d)$ denotes euclidean distance between $n$ and $d$, but we can change to a more intelligent calculation for a better result as the future work. Finally, the $Find$ identifies the destination as the next node and return the result (Lines 12-14).

We apply $PathFind$ in Figure 5. The vehicle in location $A$ plans to find a path to $F$. First it puts $B$ into set $O$ according to the direction of the vehicle in the current location, then it finds $D$ has a shorter path to $F$ compared with $C$. It chooses $D$ to start again, and finds $D$ that directly connects to $F$ and finishes. The result $A \to B \to D \to F$ is more reasonable than the shortest path $A \to E \to F$ since reversing a vehicle is rarely seen in reality. Though the delay of path $A \to B \to D \to F$ is prolonged comparing to path $A \to E \to F$, the vehicle should send more requests when it find there is no data from nearby AP. In this case, the vehicle more likely choose $B$ as its next location, then the result path is better than the shortest path since the vehicle can receive data without sending another request. In Figure 4, after installing rules at the switches along the predicted path ($A \to B \to D \to E$), the multicast address also makes $V_{2}$ receive data without sending a new request.

\subsection{Modify Address in Branching Nodes} \label{Modify address}

\begin{algorithm}[t]
\caption{ModifyAddress($s,d,path$)}
\label{Algorithm 3}
\begin{algorithmic}[1]

\STATE{Algorithm ModifyAddress($s,d,path$)}
\STATE{~~for each node $n$ in $path$ do}
\STATE{~~~~$nx$ = the next node of $n$;}
\STATE{~~~~$setRule$ = false;}
\STATE{~~~~for each rule $r$ in $n$ do}
\STATE{~~~~~~$no$ = the node connecting the output port of $r$;}
\STATE{~~~~~~if $r$ matching $s$ and $no$ != $nx$ then}
\STATE{~~~~~~~~$act$ = forward to $nx$ and modify the destination address to $d$;}
\STATE{~~~~~~~~$emitRule(matchFor(d),act)$;}
\STATE{~~~~~~~~$setRule$ = true;}
\STATE{~~~~~~else if $r$ matching $s$ and $no$ = $nx$ then}
\STATE{~~~~~~~~$setRule$ = true;}
\STATE{~~~~if $setRule$ == false then}
\STATE{~~~~~~$act$ = forward to $nx$;}
\STATE{~~~~~~$emitRule(matchFor(d),act)$;}
\STATE{~~return;}
\end{algorithmic}
\vspace{-0.06in}
\end{algorithm}

\vspace{-0.05in}

To achieve a scalable and efficient scheme for routing packets in a wired network, we propose an address modification method compared with traditional technologies unicast or multicast by leveraging the properties of OpenFlow as shown in Algorithm 2. The input $s$ implies the source of data, $d$ implies the current location of the vehicle and $path$ is computed by $PathFind$. We clarify that the rule matching $s$ means that the controller records the source address of every request, not the rule in switch matches $s$. By using our modifying address algorithm, we let packets with different source address but the same destination address match the same rule and hence decreases the number of rules and save the size of flow table (e.g., destination-driven model). Modifying the destination address of packets in branching nodes by leveraging the properties of OpenFlow is for the correctness of the algorithm. As an example shown in Figure 6, data flow $F_{1}$ and $F_{2}$ are both generated from surveillance camera $S_{1}$, and data flow $F_{3}$ comes from surveillance camera $S_{2}$. $F_{1}$ is requested from vehicle $V_{1}$. $F_{2}$, $F_{3}$ are destined to $V_{2}$. Suppose that $V_{1}$'s current location is $C$ and $V_{2}$ is at $D$, data flow $F_{1}$ and $F_{2}$ need at least two rules at switches $A$ and $B$ (and if any switches between them) in unicast, and flow $F_{2}$ and $F_{3}$ need two rules at least at switches $A$ and $B$ (and if any switches between them) in multicast, which is not efficient since $F_{1}$ and $F_{2}$ come from the same node, and $F_{2}$ and $F_{3}$ have the same destination. We modify the destination address in the header of packets in switch $B$ (as a branching node) for $F_{1}$ and $F_{2}$. By setting the destination address of packets in $F_{1}$ and $F_{2}$ ($F_{1} \to C$, $F_{2} \to D$), the number of rules in switches will be reduced and it save the size of Flow Table for more services. As an example, we assume that $V_{2}$ sends a request for data from $S_{1}$ first. $S_{1}$ sends data flow $F_{2}$ with destination address $D$ in packets header. At this moment, it only needs one rule at each switch. When $V_{2}$ asks data from $S_{2}$, the rules at switches still suit. Finally, when $V_{1}$ requires data from $S_{1}$, it just increases one rule at switch $B$ with matching $(S_{1}, D)$ with an action of modification. The header of packets can be represented by using two parameters, namely $(source address, destination address)$, so are the match conditions in Flow Table. This destination-driven mode emphasizes destination address as matching conditions in Flow Table, and generalizes the data flow demands of the same destination for reducing the size of Flow Table at switches. In Figure 4, when $V_{3}$ joins, it needs address modification at the branching switch for rule optimization.

\begin{figure} [t]
\begin{center}
\includegraphics[width=0.9\columnwidth]{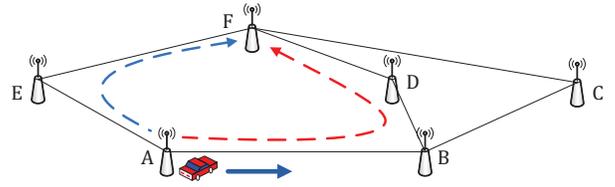}

\caption{The blue dash line is the shortest path between the current location and destination, while the red dash line is a more likely path that the vehicle may choose according to the direction.} \label{fig5}
\end{center}
\vspace{-0.2in}
\end{figure}

\begin{figure} [t]
\begin{center}
\includegraphics[width=1\columnwidth]{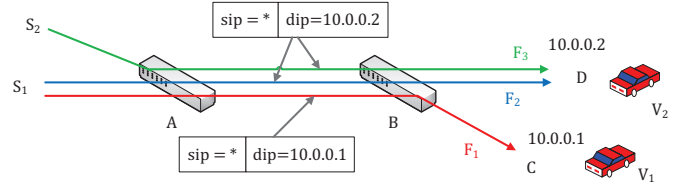}
\caption{Data flow $F_{1}$ (red line) comes from $S_{1}$ and its destination is $C$; Data flow $F_{2}$ (blue line) comes from $S_{1}$ and its destination is $D$; Data flow $F_{3}$ (green line) comes from $S_{2}$ and its destination is $D$. The matching conditions only depend on the destination address.} \label{fig6}
\end{center}
\vspace{-0.3in}
\end{figure}

\begin{figure*} [t]
\begin{center}
\begin{tabular}{cc}
\includegraphics[width=1.9\columnwidth]{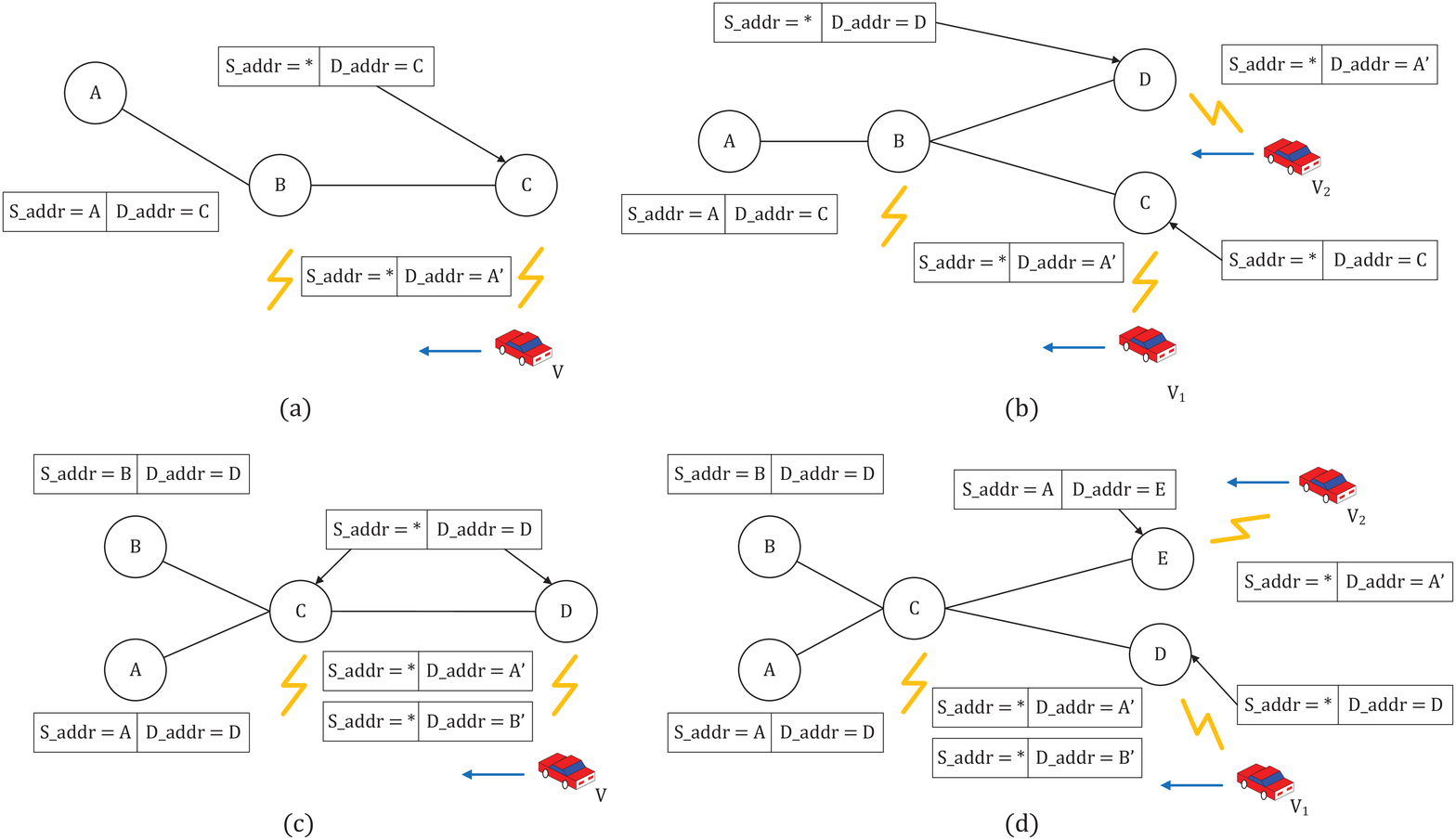}
\end{tabular}
\caption{Subfigure (a) shows a \textit{1 to 1} pattern that $V$ wants to receive data from $A$; (b)shows a \textit{N to 1} pattern that $V_{1}$ and $V_{2}$ want to receive data from $A$; (c)shows a \textit{1 to N} pattern that $V$ wants to receive data from $A$ and $B$ at the same time; (d) shows a \textit{N to N} pattern that $V_{1}$ wants to receive data from $A$ and $B$ while $V_{2}$ wants to receive data from $A$.} \label{fig7}
  \end{center}
  \vspace{-0.3in}
\end{figure*}

\subsection{Examples} \label{Examples}

In this section, we will give four examples to show how our scheme works and then describe the details about rules installed at switches. By analyzing the process of data transmission in real-time query service, we summarize four patterns as shown in Figure 7.

\subsubsection{1 to 1 \rm{case}} \label{1 to 1}

In the simplest situation, there is only one vehicle asking for real-time query service from one surveillance camera. As shown in Figure 7 (a), vehicle $V$ sends a request to camera $A$ and then receives data flow coming from $A$. Therefore, the packets generated from $A$ is $(A, C)$, and match conditions at $B$,$C$ are both $(*, C)$. The packets multicast at $C$ is $(*, A^{'})$ where $A^{'}$ is multicast address of $A$. In this scenario, the path found by $PathFind(s,d,v)$ is $C\to B \to A$, which is the most likely path that the driver would choose. Then we install rules for multicast with $(*,A^{'})$ at $B$ previously. When the vehicle arrives to $B$, it can receive data without interruption.

\subsubsection{N to 1 \rm{case}} \label{N to 1}

When a camera connecting to multiple vehicles, there needs to modify the destination address at the branching nodes (like $B$ in Figure 7 (b)). The cost of modification would not be large since it is a necessary work to copy the packet, and the address for modifying can be preserved at switches. The buffer inconsistency problem can not happen since modification is carried out at switches, and the rules must be installed before data flows arriving. At the beginning, $A$ sends packets $(A, C)$ since vehicle $V_{1}$ at $C$ requests faster than $V_{2}$ at $D$. When $V_{2}$ sends a request, the controller (not shown in this figure) installs a new rule with a modification action (can be implemented in OpenFlow). As a result, switch $B$ forwards two packets $(A, C)$ and $(A, D)$ to different ports. The data flows forwarded (by AP) at $B$, $C$ and $D$ are all $(*, A^{'})$.

\subsubsection{1 to N \rm{case}} \label{1 to N}

A vehicle may connect to multiple cameras simultaneously. As descried in Figure 7 (c), vehicle $V$ wants to receive data from $A$ and $B$ at the same time. The packets generated from $A$ and $B$ are $(A, D)$ and $(B, D)$. At switches $C$ and $D$, the matching conditions are both $(*, D)$. There only needs $one$ rule to meet the requirements for different packets. When the packets $(A, D)$ come to $C$, the rule $(*,D)$ matches the packets and then forwards to $D$. The switch $D$ will follow the same procedure as $C$. $C$ and $D$ also need to change the destination address of the packets for multicast. The packets for multicast at $C$ and $D$ are $(*, A^{'})$ and $(*,B^{'})$, where $A^{'}$ and $B^{'}$ are the multicast address.

\subsubsection{N to N \rm{case}} \label{N to N}

We combine the \textit{1 to N} and \textit{N to 1} for a more common scenario \textit{N to N} in Figure 7 (d). Vehicle $V_{1}$ requests real-time data from $A$ and $B$, and $V_{2}$ requests data from $A$ only. There needs a modification at $C$ just for packets matching $(A, *)$, since $C$ is a branching node when data source $A$ transfers packets to $D$ and $E$ at the same time. The packets coming from $B$ are just forwarded to the given port without any modification.

The \textit{N to N} scenario is a general case that give details about how to implement rule optimization. Furthermore, we can observe that the theoretical upper bound of the table size of the proposed rule optimization approach is related with the number of devices which the driver is connecting to at the same time. $N_{d}$ denotes the number of devices, $R_{trad}$ denotes the number of rules for the traditional approach and $R_{opt}$ denotes the number of rules for the proposed rule optimization approach, then we have $R_{opt} = R_{trad}/N_{d}$ is the upper bound of the table size we save. In this case, there are all devices connecting to the same switches.

%
%

\section{Performance Evaluation} \label{Evaluation}

\subsection{Testbed}

We use Floodlight\cite{Floodlight} as the controller and Mininet\cite{mininet} to build a SDN environment to testify rule optimization strategy in SDIV. We run Floodlight on a server, with 16 AMD Opteron(tm) processor 6172 and 16G memory. Our server software includes Linux kernel version 2.6.32. We run Mininet on a separate server and servers are connected by a 10Gbps Ethernet network.

\subsection{Effects of Rules Optimization}

We use the real data of traveling trace in Shanghai \cite{shanghai} as a common scenario to show the benefit of rule optimization. Figure 8 (a) shows a snapshot around People's Square in Shanghai. Traveling traces of vehicles are composed of GPS data at different times. At the first time ($t_{1}$), there are only five vehicles in this area. At $t_{2}$, there are another two vehicles join. At $t_{3}$, vehicle $V_{6}$ appears in the area. The appearance time and disappear time of vehicles are illustrated in Figure 8 (b). Interval time between any two moments is 120 seconds. We make any mark belong to the same moment if the difference between their timestamps and the moments is shorter than 30 seconds. Although these GPS marks at the same moment may not have exactly the same timestamp, it can roughly say that these vehicles move to these locations very closely at that moment by limiting the difference between the timestamp of GPS marks and the time of moments to a certain range. In this case, we set it to 30 seconds to ensure that every GPS mark can reflect the real location at that moment.

Figure 8 (c) is a sketch map of Figure 8 (a). Arrow lines with different colors illustrate the directions of vehicles according to the order of these timestamps in GPS marks as shown in Figure 8 (a). We assume that there is always one road-side AP (switches in Figure 8 (c)) around the GPS marks for data transmission available at any time. We make every vehicle in sketch map have a data transfer demand at the first moment they appear. Therefore, $V_{1}$ and $V_{2}$ near the intersection of line 8 and line 2 want to receive data from both $B$ and $C$, and also have the same path along line 8. $V_{3}$ locating on the convergence of Beijing E Rd and Fujian Middle Rd also has demands of receiving data from $B$ and $C$. $V_{4}$, $V_{5}$ and $V_{6}$ have the same target $A$ but choose different paths. $V_{7}$ asks the data from $D$. $V_{8}$ has the target $B$ only. All these vehicles require a persistent data flow service until they move to their destinations. Then we need to install rules at every necessary switches for their requirements. Simple installation which we compared with is a general method forwarding packets to given ports based on the source address. By rule optimization, it will merge the rules which have the same destination (destination-driven mode) in \textit{1 to N} pattern like $V_{3}$ connecting to $B$ and $C$, and modify headers of packets at branching nodes in \textit{N to 1} pattern. The result is shown in Figure 8 (d). The decreasing at $t_{4}$ is because of the timeout in rules we set. Figure 8 (e) shows the number of vehicles at different moments. Figure 8 (f) shows the delay time of different vehicles in two strategies. The delay time we evaluate here is the longest one if a vehicle connects two cameras at the same time (we choose $C$ instead of $B$ for evaluating $V_{1}$'s delay time). And we can see from Figure 8 (f) that our rule optimization strategy hardly influences the performance of data transmission though it needs address modification in the header of packets. By rule optimization, it reduces the number of rules and saves the space at Flow Table for scalability.

\begin{figure} [t]
\begin{center}
\begin{tabular}{c}
\includegraphics[width=0.85\columnwidth]{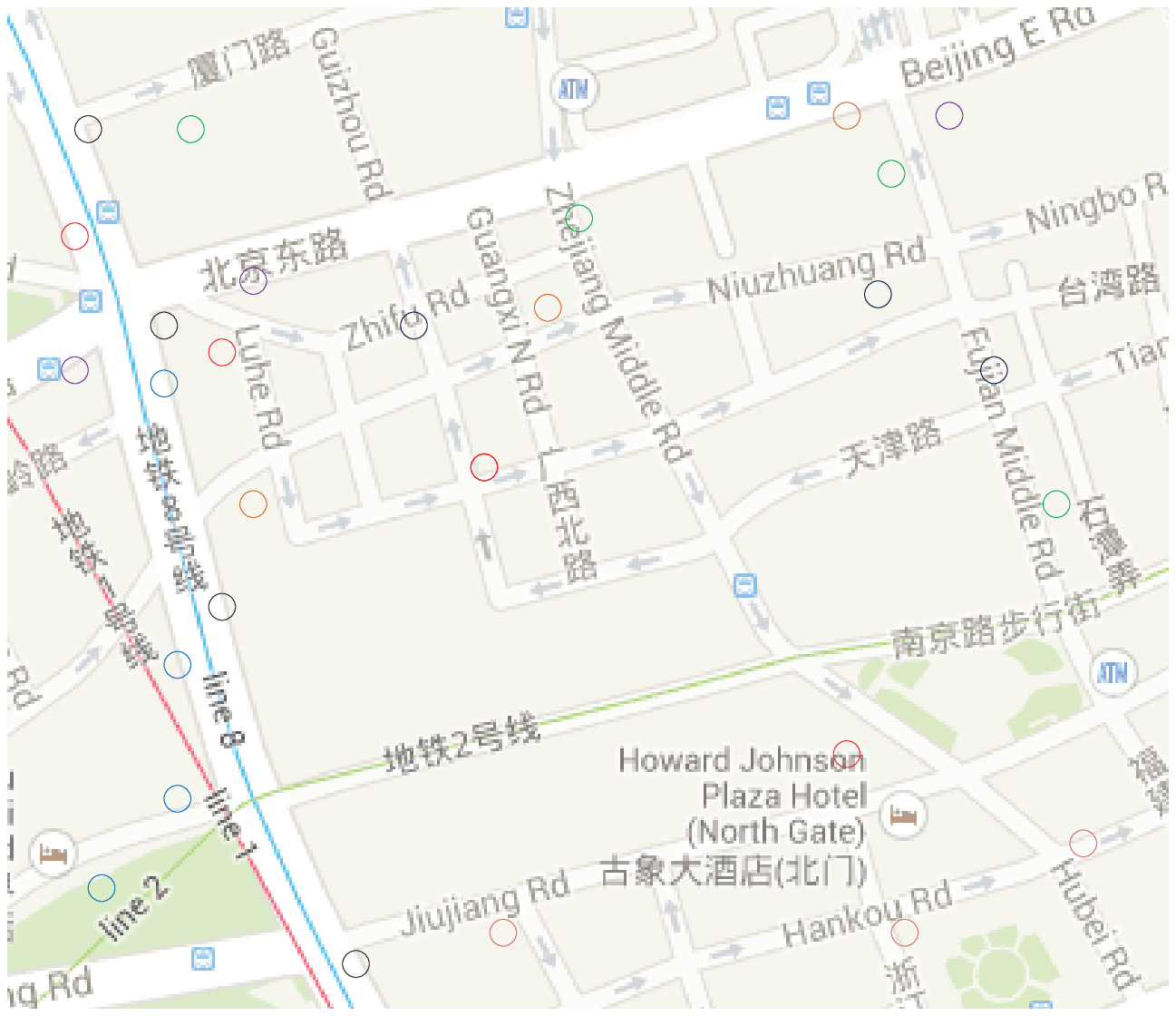} \\
(a)
\end{tabular}
\begin{tabular}{cc}
\includegraphics[width=0.45\columnwidth]{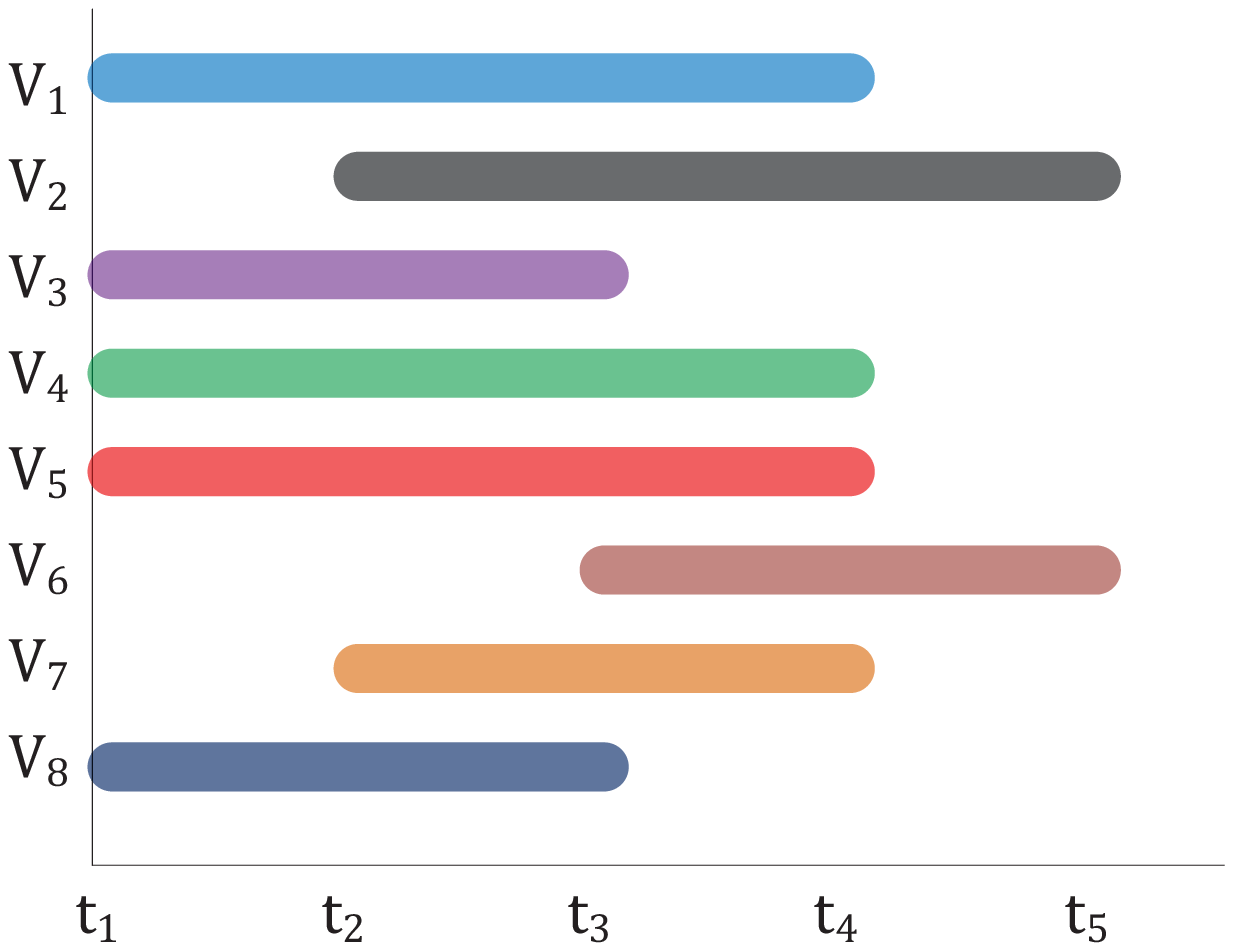}&
\hspace {-0.1in}
\includegraphics[width=0.45\columnwidth]{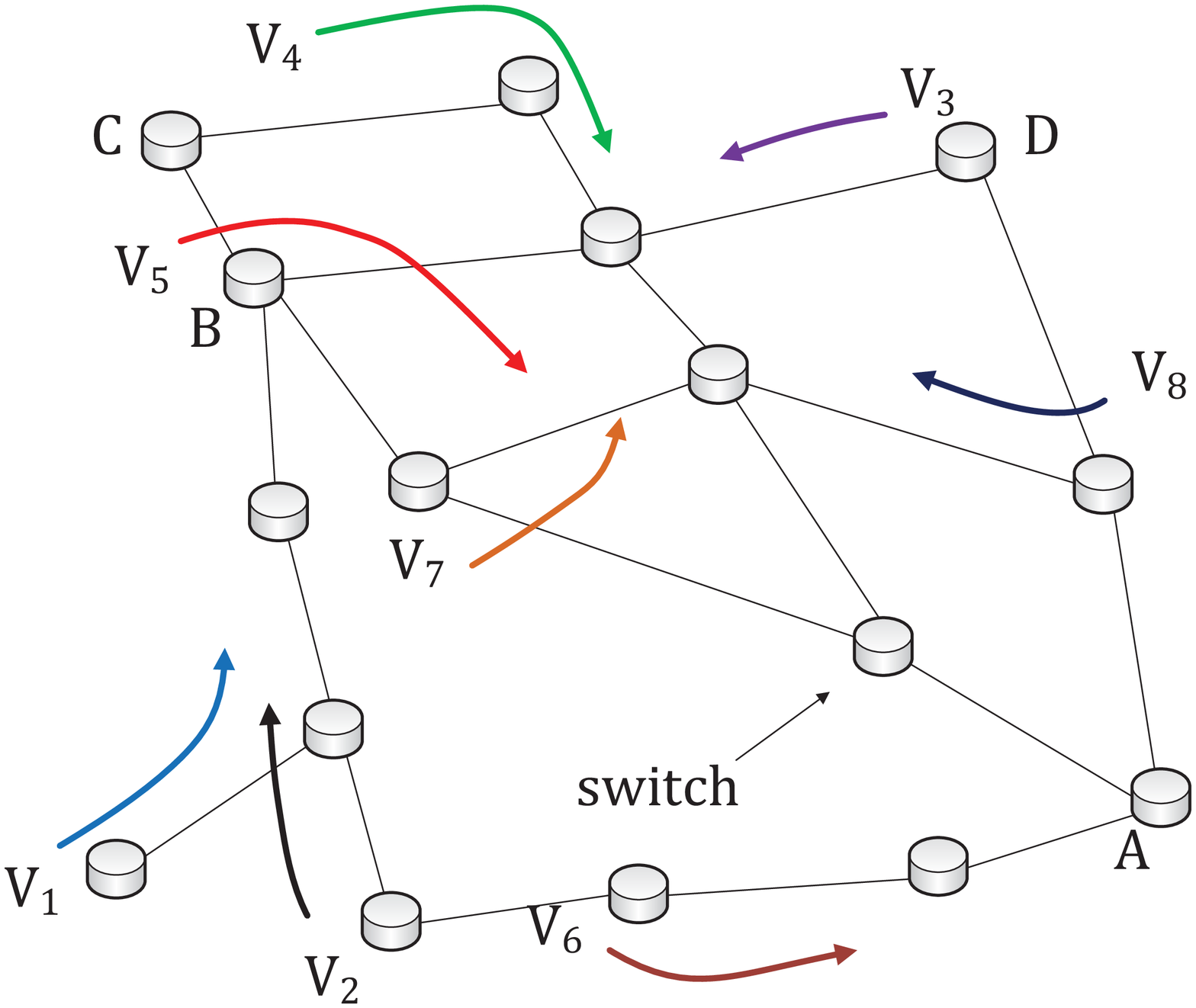} \\
(b) & ~~~(c)
\end{tabular}
\begin{tabular}{cc}
\includegraphics[width=0.5\columnwidth]{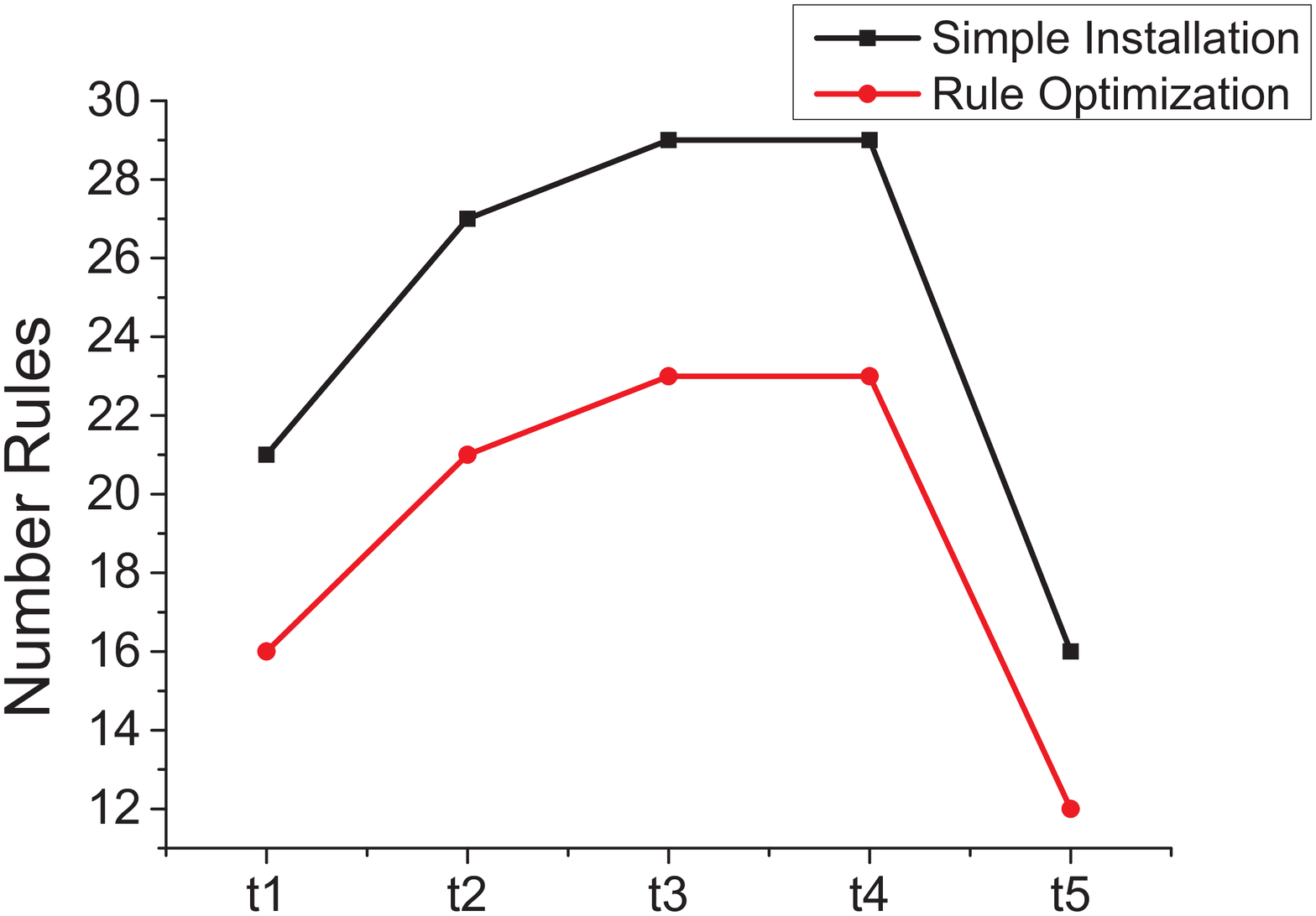}&
\hspace {-0.2in}
\includegraphics[width=0.5\columnwidth]{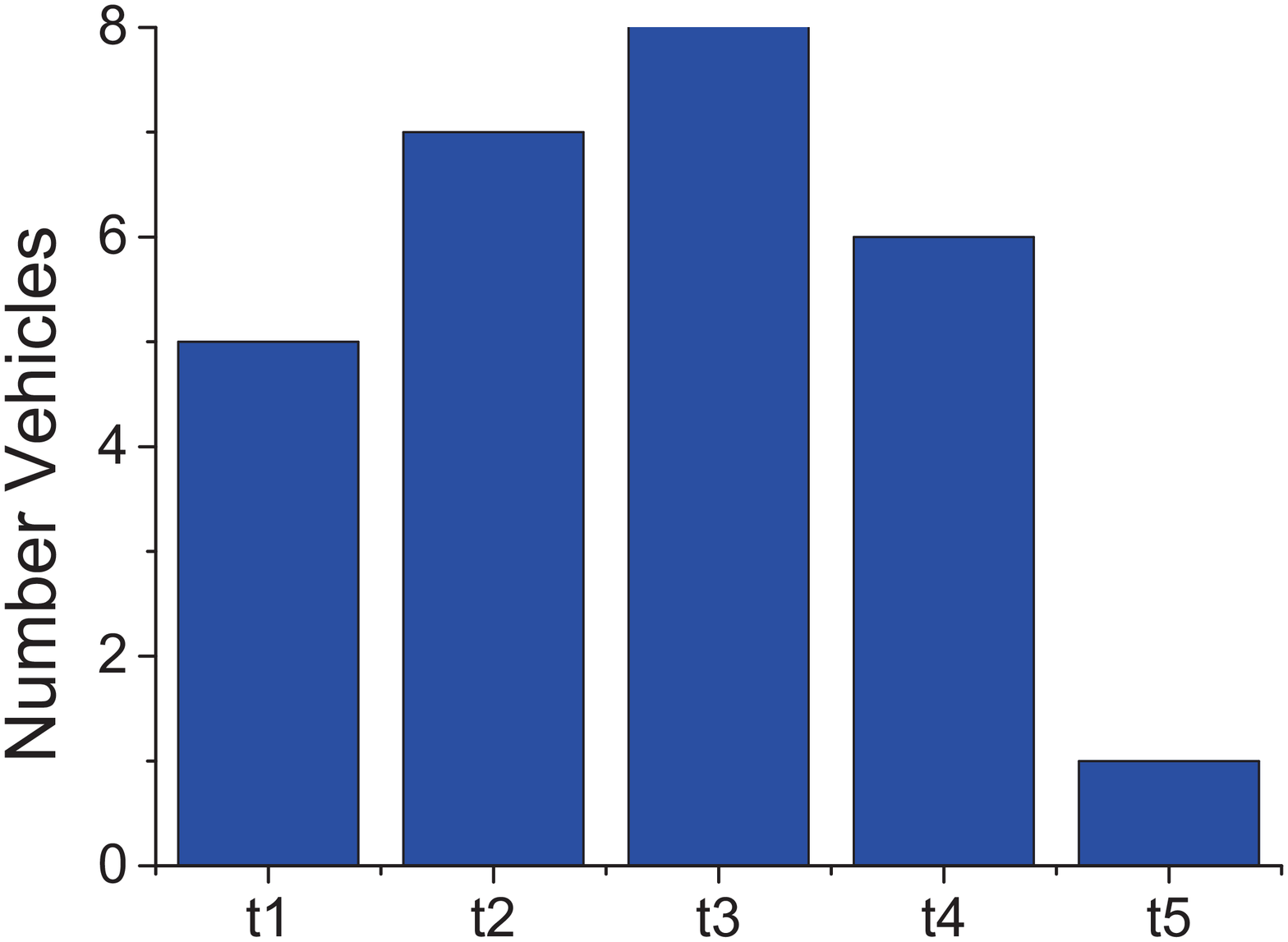} \\
(d) & (e)~~~
\end{tabular}
\begin{tabular}{c}
\includegraphics[width=0.9\columnwidth]{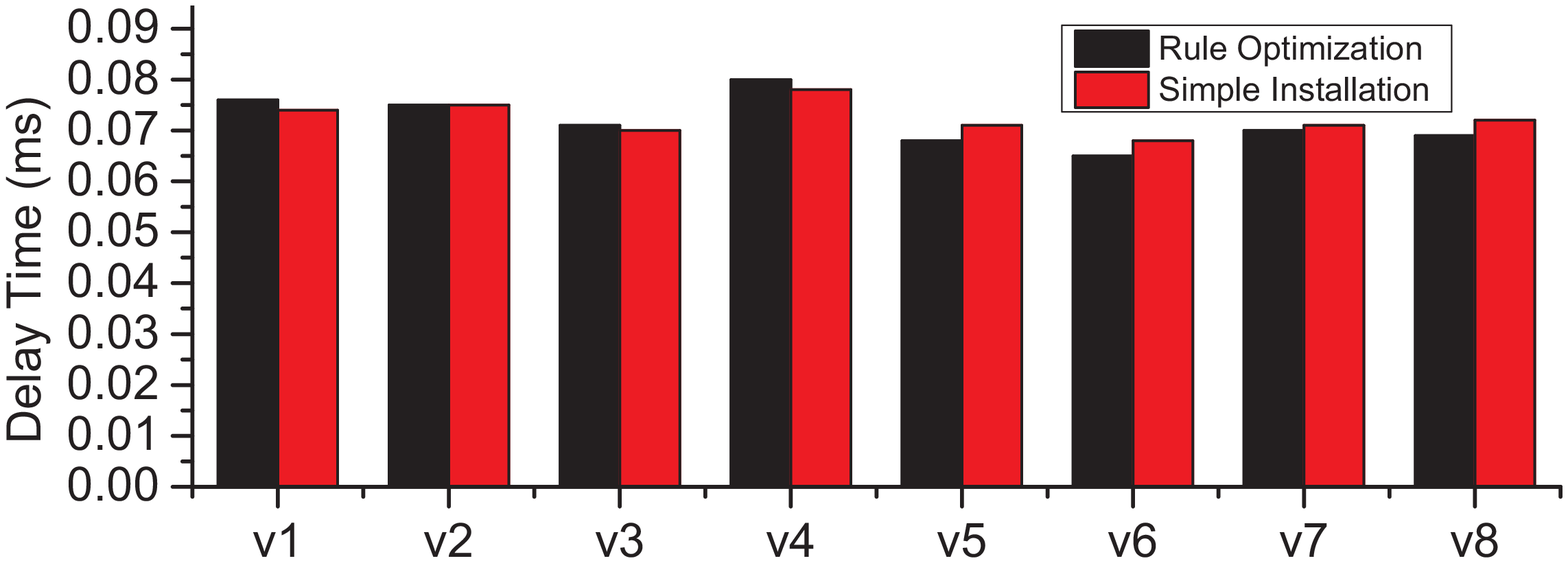} \\
(f)
\end{tabular}
\caption{Subfigure (a) shows a snapshot around People's Square in Shanghai; (b) shows the appear and disappear time of vehicles; (c) is a sketch map of (a) with the direction of vehicles; (d) shows different number of rules for two strategies; (e) shows the number of vehicles at different moments; (f) shows the delay time of vehicles in two strategies.} \label{fig8}
  \end{center}
  \vspace{-0.35in}
\end{figure}

\subsection{Analysis}

We study the two patterns (\textit{1 to N} and \textit{N to 1}) in details and show how they influence results differently. We compare the number of rules and the delay time with simple installation to show that it does not affect the performance of data transmission. Figure 9 (a) shows \textit{N to 1} mode that there are multiple vehicles connecting to one surveillance camera ($B$). We pick $A$ which has the longest path to evaluate delay time. It means every packet transferred from $B$ to $A$ will modify the headers (destination IP address) at each switch and then forward to the next switch. And we compare our method with the simple installing that every packet just forwards to given ports without modifying the header. Figure 9 (b) shows that, as the switch number increasing, delay time of both methods get large but have little difference between each other. The packet modification process in rule optimization strategy barely influences the performance. Figure 9 (c) shows the number of rules at switches. The black line represents simple installing mode that node $B$ transfers data flow across networks to every node at the same time. Therefore, there only needs one rule at each switch with $(*, B^{'})$ and $B^{'}$ means the multicast address. The red line depicts rule optimization strategy that we modify the destination address at every branching node and then forward the packets to given ports. There also needs only one rule at each switch comparing to simple installing mode.

Figure 9 (d) shows \textit{1 to N} patten that the vehicle $A$ connects to multiple cameras ($B, C, D$) at the same time. Figure 9 (e) shows the delay time according to the different number of switches. Without rule optimization, there need three rules for each source node at each switch, $(B, *), (C, *), (D, *)$, which is a source-driven mode (forwards packets based on data source). In our rule optimization strategy, there only needs one rule, $(*, A)$, at each switch and it still support \textit{N to 1} pattern as discussed above. Figure 9 (f) shows the number of rules and our method has a more compact rule table than simple installing mode.

We draw two conclusions from the result. First, the modification of packets at branching nodes has little impact on the performance of data transmission. Second, for the best case of rule optimization (\textit{1 to N}), the number of reduced rules is proportional to the number of data sources, and even for the worst case of rule optimization (\textit{N to 1}), the number of rules at switches is equal to that of simple installing mode which just forwards packets to different ports without any strategy.

\begin{figure*} [t]
\begin{center}
\scalebox{0.9}
{
\begin{tabular}{ccc}
\includegraphics[width=0.65\columnwidth]{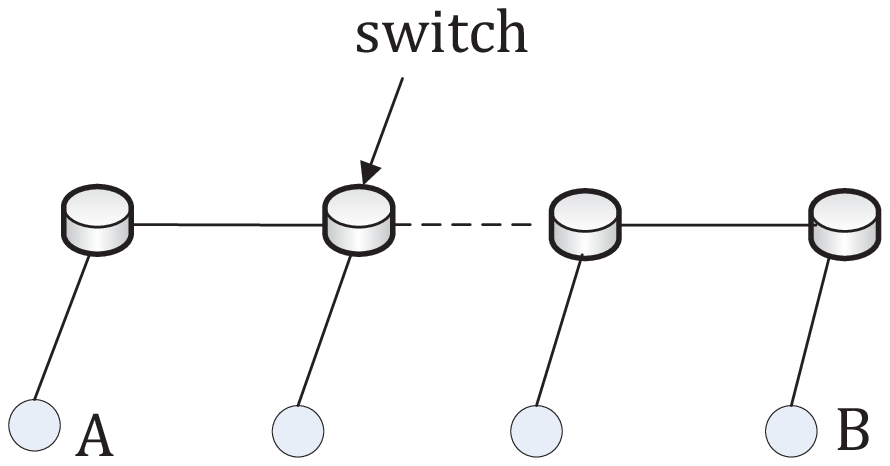}&
\includegraphics[width=0.65\columnwidth]{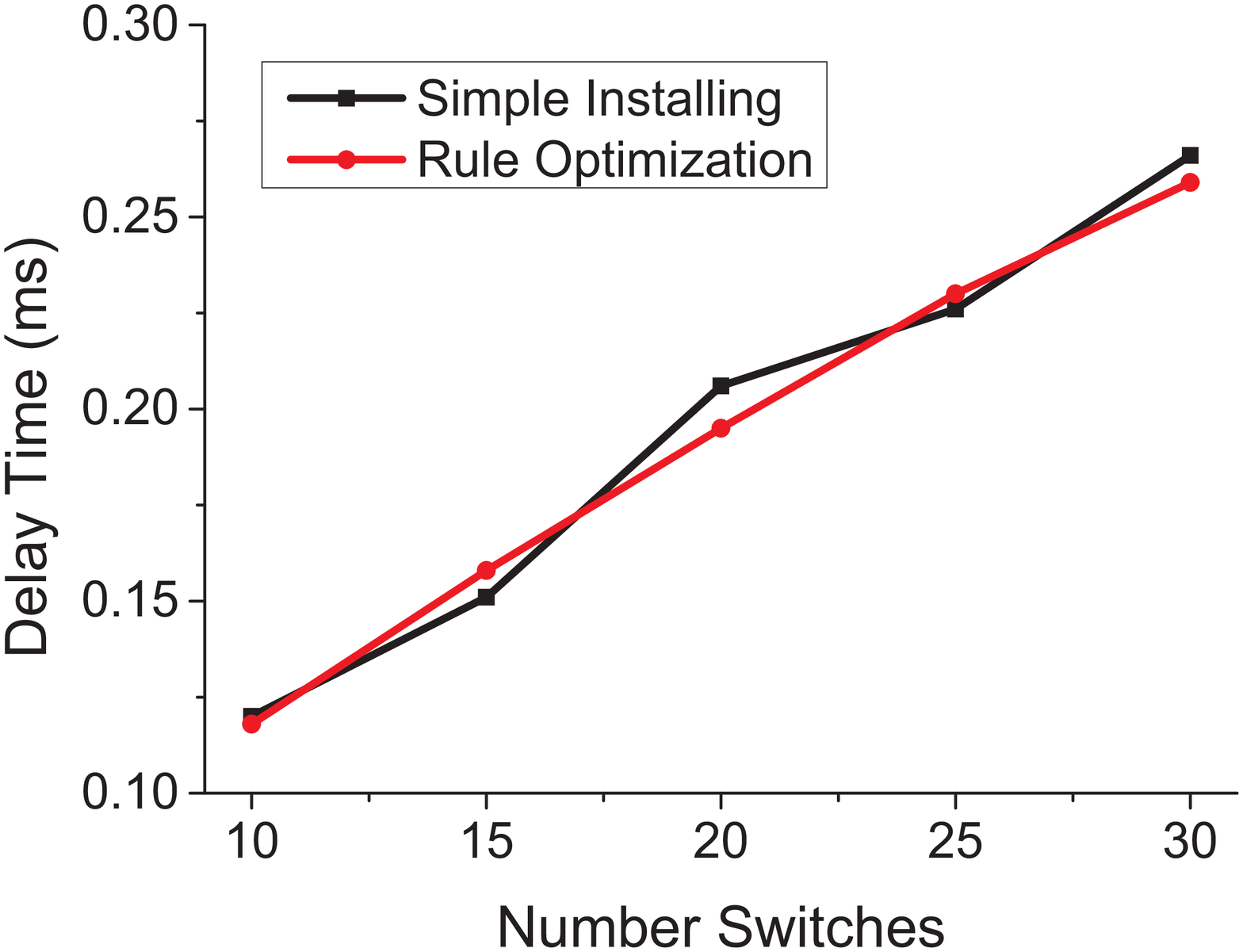}&\hspace{-0.1\columnwidth}
\includegraphics[width=0.65\columnwidth]{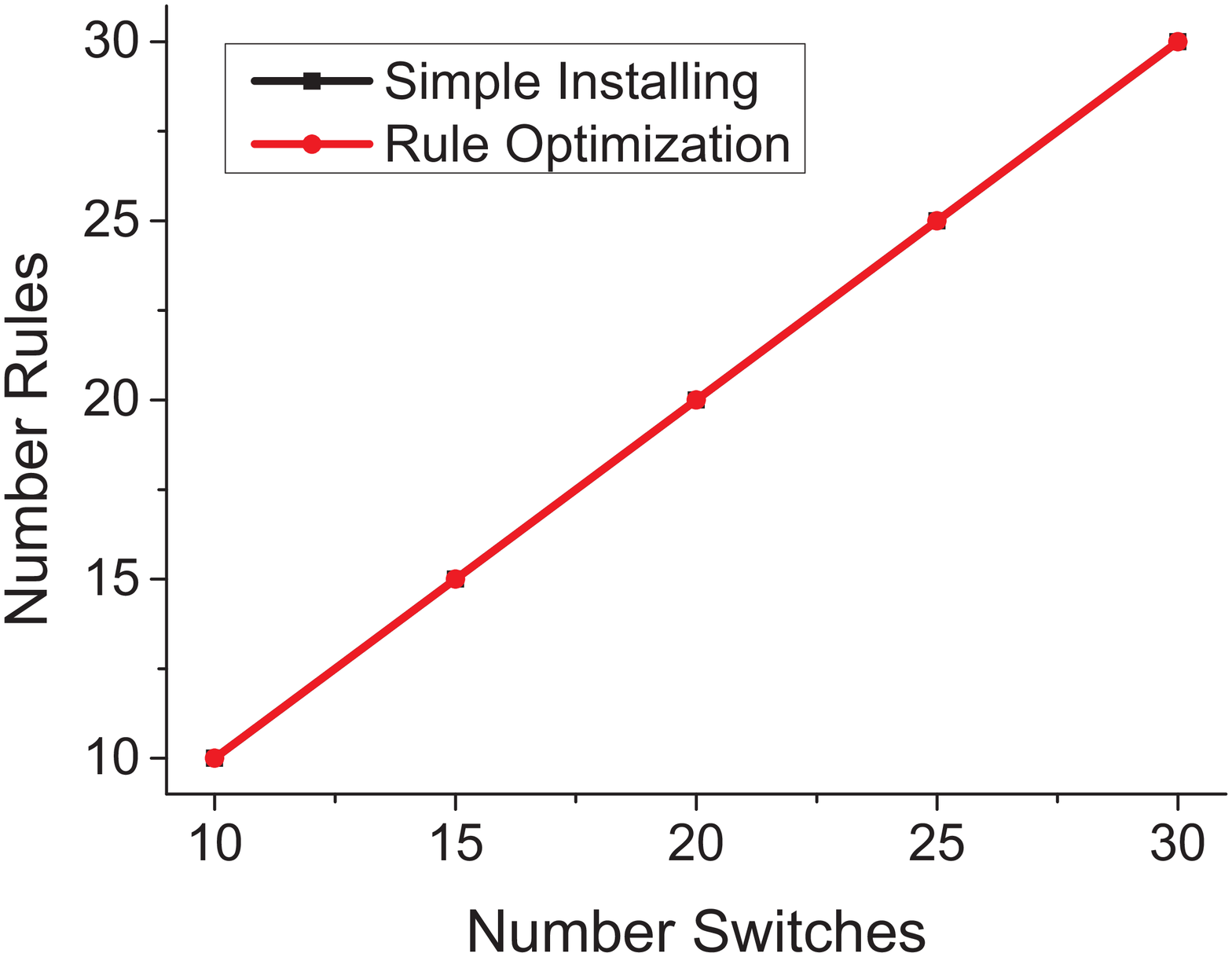} \\
(a) & (b) & (c)
\\
\includegraphics[width=0.65\columnwidth]{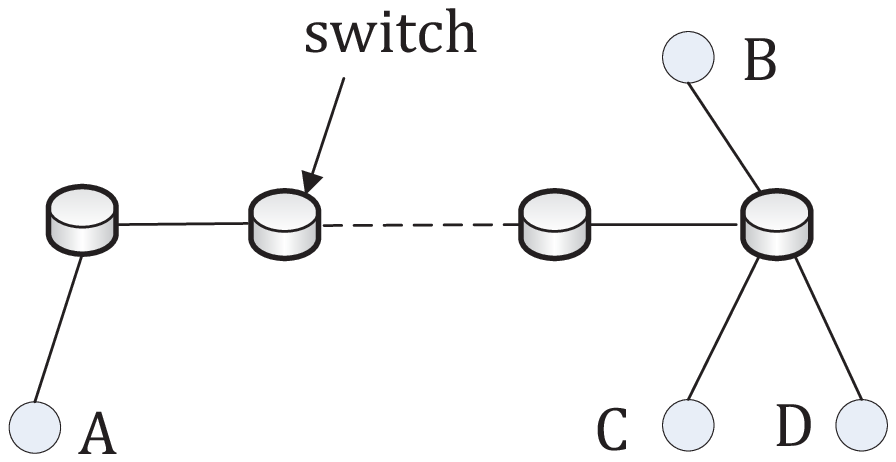}&
\includegraphics[width=0.65\columnwidth]{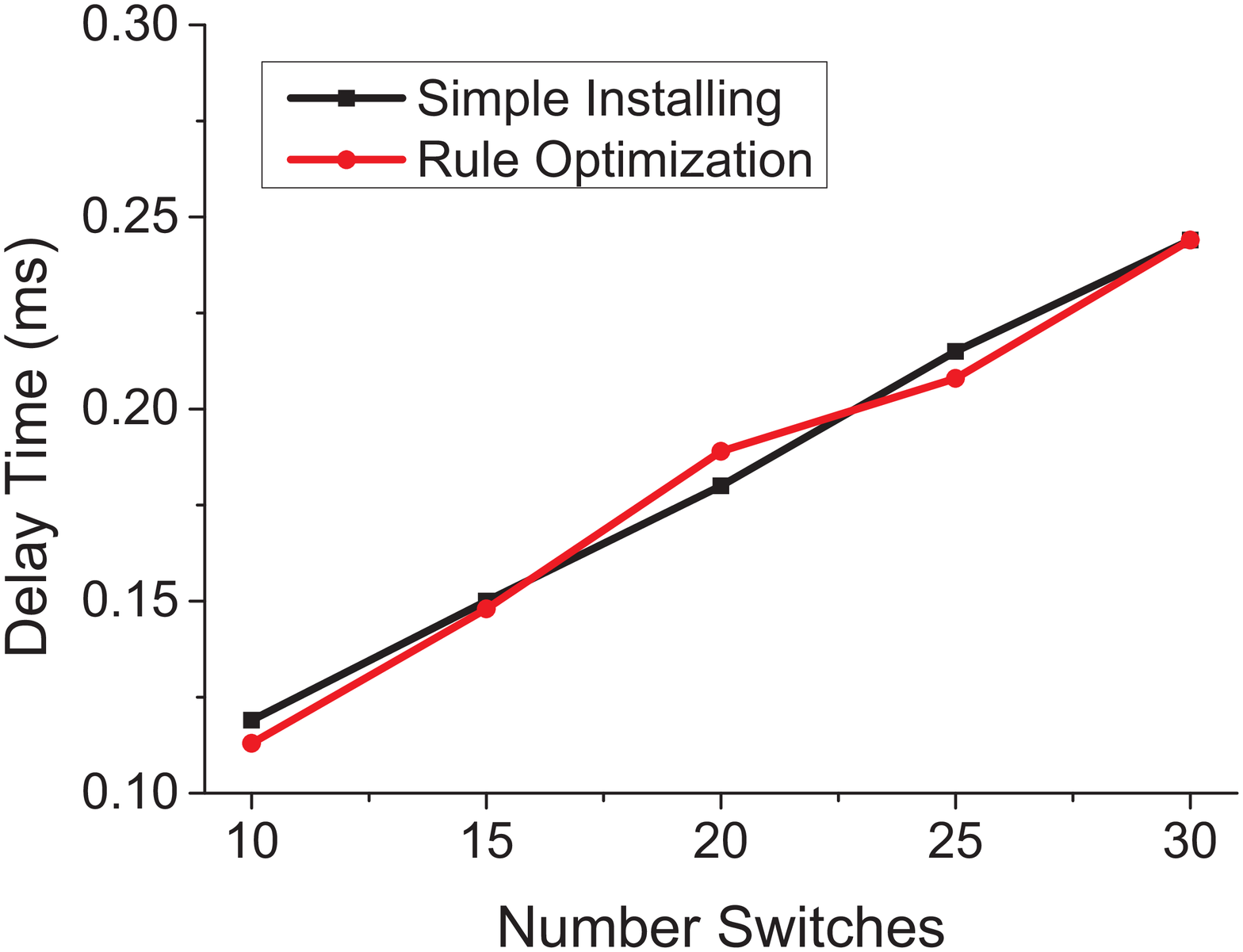}&\hspace{-0.1\columnwidth}
\includegraphics[width=0.65\columnwidth]{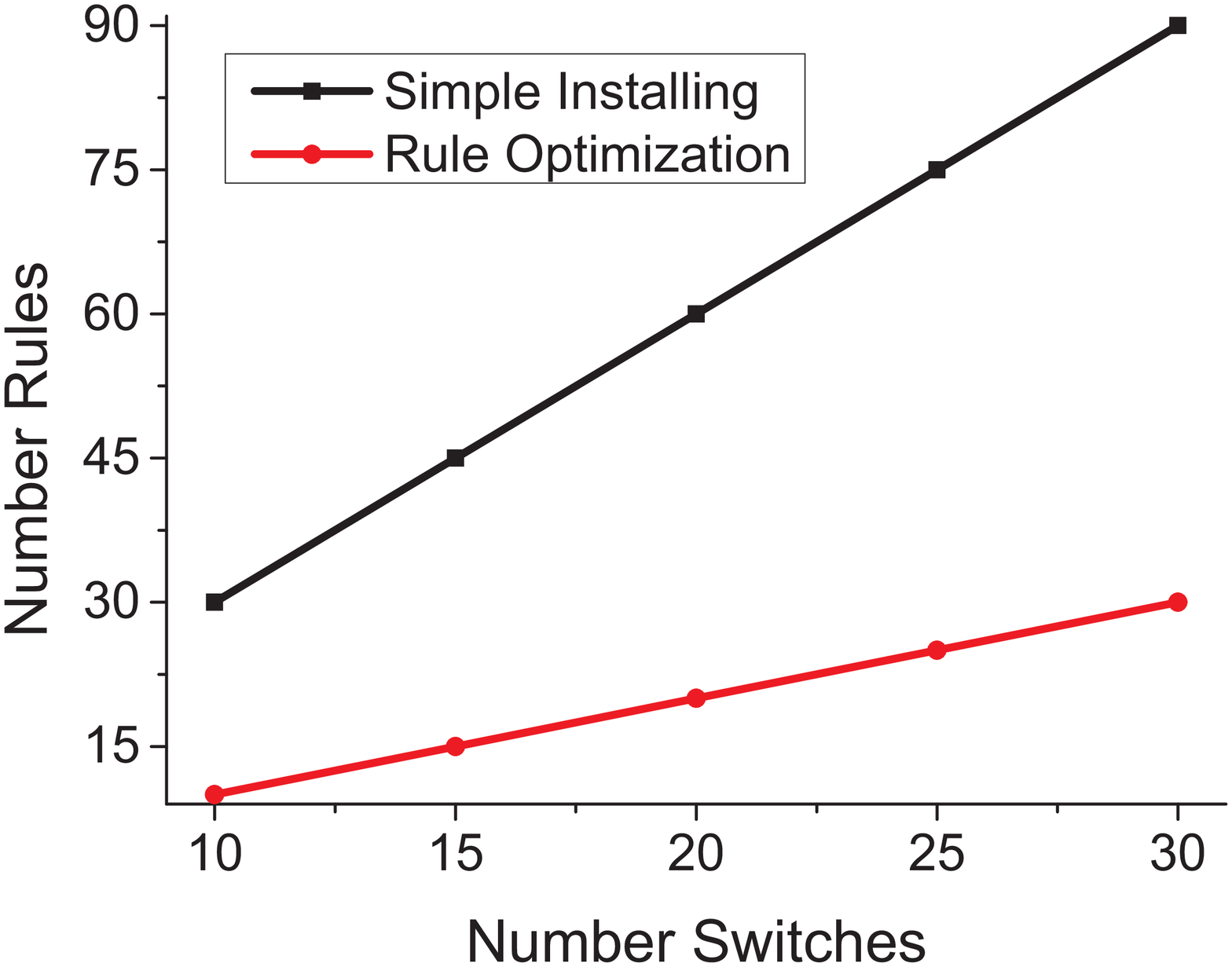} \\
(d) & (e) & (f)
\end{tabular}
}
\caption{Subfigure (a) shows a topology of \textit{N to 1} mode; (b) shows delay time on different number of switches in (a); (c) shows number rules on different number of switches in (a); (d) shows a topology of \textit{1 to N} mode; (e) shows delay time on different number of switches in (d); (f) shows number rules on different number of switches in (d)} \label{fig9}
  \end{center}
\vspace{-0.3in}
\end{figure*}

\section{Related Work} \label{Related Work}

To the best of our knowledge, SDIV is the first architecture that combines IoV with SDN by using a centralized controller to manage the network devices. Hence, there is rarely related research work and most we list are either related to vehicular networks or SDN. Most applications and services (road security, fleet management, navigation, billing, multimedia, etc) in IoV rely on data exchanged between the vehicle and the roadside infrastructure (V2I) and between vehicles (V2V) \cite{ernst2006information}.
Several research works \cite{ott2005disconnection}\cite{balasubramanian2008interactive} have demonstrated the feasibility of providing connectivity via road-side APs and the ubiquity of WiFi.The studies \cite{joshi2011vehicular}\cite{chen2009mobtorrent}\cite{xu2011utilizing} deal with the efficient communication between APs and vehicles. In \cite{xing2013approaching}, the authors propose a novel system design and implementation for realtime reliable roadway communications, for providing safety messages to users in a realtime and reliable manner. For routing and data forwarding, there are \cite{li2013horizon}\cite{zhang2014geomob}\cite{leontiadis2010extending}\cite{zhu2013zoom} that study the routing protocol and fast forwarding. To manage vehicular networks, a centralized policy framework is introduced in \cite{hare2012policy}, Virtuoso, that manages spectrum resources while ensuring users have suitable access for their communication needs. In \cite{malandrino2011content}, the authors focus on content downloading in vehicular networks. Paper \cite{ahn2012risa} presents the Road Information Sharing Architecture (RISA), the first distributed approach to road condition detection and dissemination for vehicular networks.

To establish a new architecture in vehicular networks, Named Data Networking is applied to networking vehicles and enables networking among all computing devices independent from whether they are connected through wired infrastructure, ad hoc, or intermittent Delay Tolerant Network in \cite{grassi2014vanet}. In \cite{yap2010blueprint}, the authors present an OpenFlow \cite{mckeown2008openflow} Wireless networks to achieve a free travel between any wireless infrastructures by separating the network service from the underlying physical infrastructure.

In the area of software-defined network, SDN is applied to improve network management in \cite{kim2013improving}. For rule optimization, a distribution framework is proposed for decomposing large switch tables into small ones to address limited size of switch tables \cite{kanizo2013palette}. In \cite{voellmy2013maple}, the authors introduce Maple to discover reusable forwarding decisions and reduce the number of rules by a trace tree structure that records access on a specific packet. A centralized traffic engineering with SDN has been proposed in \cite{hong2013achieving} and \cite{jain2013b4} with the idea of classifying the services based on their performance requirements.

\section{Conclusion and Future Work} \label{Conclusion}
This paper proposes a new architecture SDIV to address the issues of the proprietary and closed way of operating hardwares in network equipments, which slows down the progress of new services deployment and extension in IoV. But, simple installation of rules is not efficient since the characteristics of IoV. Rule optimization is necessary in SDIV. Our rule optimization method can reduce the size of Flow Table but not degrade the performance of data transmission. The separation of wired data plane from wireless data plane and the destination-drive mode are suited well for the characteristic of SDIV. Evaluation shows that our rule optimization strategy reduces the number of rules without losing the performance of data transmission.

In our future works, we plan to address flowing issues:

(1) We assume that there is no limitation in the capacity of switches, which means that the performance of data transmission does not degrade as the number of vehicles connecting to one camera increasing. Actually, it is necessary to design an allocation mechanisms for controlling data flows based on the whole network status.

(2) The controller should gather geographic information around the switches (APs) for making further decisions. This is an important work in reality. We will consider an efficient interface such as letting switches uploading their conditions at regular intervals.

(3) We only considered real-time query service in this work. As a future work, it is necessary to investigate data uploading service and achieve a coordinator between the different services based on the performance demands. This includes identifying the conditions of switches and the service types of data flows.

\bibliographystyle{IEEEtran}
\bibliography{bibsdiv}

\end{document}